# "Is Our Organization *Actually* Measuring Productivity?"

*How contrasting organizational and individual measures of engineering success is an opportunity to drive engineering transformation*

— Carol S. Lee [1], Morgan Ramsey [1], & Catherine M. Hicks [1,2]

*Abstract -* **Research problem:** *Defining developer productivity is both challenging and top-of-mind for engineering leaders. However, engineering teams continue to struggle to define and measure it.* **Research questions:** *Which metrics are Organizations prioritizing on productivity, and do their engineers agree with this? Are there differences in currently used versus preferred metrics between engineering leaders, managers, and individual contributors?* **Literature review:** *Previous research highlights that common misconceptions about developer productivity lead to harmful and inaccurate evaluations of software work, pointing to the need for organizations to differentiate between measures of production, productivity, and performance as an important step that helps to suggest improvements to how we measure the success of engineering teams.* **Methodology:** *Using a card sort, we explored how a Three Layer Productivity Framework was used by 16 software engineers at a Software Engineering-focused conference to rank measures of success, first in their organizations' current practice and second in their individual beliefs about the best ways to measure engineering success. Our participants represented various backgrounds, including engineering leaders and managers as well as individual contributors.* **Results and discussion:** *Overall, participants preferred organizations to 1)* **continue** *their prioritized focus on performance layer metrics, 2)* **increase** *the focus on productivity metrics and 3)* **decrease** *their focus on production metrics. When asked about their organization's current metrics, while all roles reported a current focus on performance metrics, only ICs reported a strong focus on production metrics. When asked about metrics they would prefer, all roles preferred more performance metrics but only leaders and ICs also wanted productivity metrics. While all participants were aligned on performance metrics being a top preference, there was misalignment on which specific metrics are used. Our findings show that when measuring developers' success, organizations should continue measurement using performance metrics, consider an increased focus on productivity metrics, and consider a decreased focus on production metrics.*

---

1. [Developer Success Lab](#), Pluralsight Flow
2. Email: flow-research@pluralsight.com

**INTRODUCTION**
Despite a longstanding emphasis on developer productivity, engineering teams continue to struggle to define and measure it. This may be in part due to two misconception "traps" in the way we conceptualize productivity: 1) a conflation between related constructs like production and performance, leading to organizations making conclusions from the wrong metrics [1], and 2) a lack of including context-dependent and multidimensional measures over time, leading to organizations overlooking how definitions of success might change depending on the constraints at hand and when we should take a longer perspective over time [2, 3].

Because of the difficulty in thoughtfully defining productivity, software organizations have largely failed to come to a shared understanding of it and failed to "translate" productivity in engineering work throughout other parts of the business [4]. These inconsistent practices have subsequently led to software teams essentially reinventing their own definitions of productivity, much of which remains unspoken and unshared between team members, and between individual contributors (ICs) and managers [2]. Where organizational definitions of productivity *are* pursued, further research suggests that developers' own reflections and insights about productivity are hardly, if ever, surfaced in organizational discussions of software velocity and work [5]. Despite the importance of developer productivity, engineering leaders have also continued to struggle to set appropriate "productivity" targets within their organizations.

This incoherence may lead to multiple complex outcomes that appear different on the surface, but spring from the productivity misconception "traps" described above. For example: 1) a "maintenance" software team making iterative improvements to a legacy system is unfairly seen as unproductive relative to a fast-paced "feature" delivery team 2) a highly valued senior tech lead is fired during a broad-sweeping layoff, because their productivity was represented in architecture, mentorship, and other collaborative activities rather than individual "production" software metrics 3) an engineering organization finds itself struggling with a constant "boom and bust" cycle, where coding activities seems to spring forward but then decline precipitously, because software teams go through cycles of trying to "perform to the metric" and push up activity measures during moments when they feel their work is under scrutiny.

**The Three Layer Productivity Framework**
How can engineering organizations and leaders move past the historical "traps" of misunderstanding developer productivity? To guide engineering managers and leaders, who may wish to introduce new metrics, adapt existing measures, or improve the shared understanding of organizational goals within and between software teams, we previously described a conceptual framework breaking down the three "layers" of productivity [6]. This framework synthesizes the research across the social and clinical sciences and software engineering to distinguish the differences between **production**, **productivity**, and **performance** (see Figure 1).

The Three Layer Productivity Framework [6] draws from research-backed models of human productivity to highlight three levels for characterizing developers' work, in increasing order of maturity and accuracy. The first layer, **production**, refers to the amount of output, regardless of resources provided and used [7, 8]. In the context of engineering teams, this may be captured by metrics such as raw counts of merged pull requests, number of code comments, or quantity of reviews done. Although production metrics may be able to provide information about large trends over time within a single team (e.g., a team merging zero pull requests in the past six weeks could be a red flag about a serious issue), they cannot provide insight into productivity. Critically, "production" measures are unlikely to be an accurate basis for making comparisons between teams.

The second layer is **productivity**, which refers to the quantity of output, *given the resources provided* [3]. Because productivity fully depends on the resources used and provided, production could be identical across two teams, while productivity vastly differs. To track productivity, production could be considered in conjunction with an assessment of resources, team size, and bandwidth to understand rates of production for different contexts. This may allow for a more accurate picture of software teams, as well as begin to create an opportunity for leaders to make comparisons between teams. However, productivity measures may be an inaccurate basis for making conclusions about long-term impact.

The final layer is **performance**, which refers to factors such as the flexibility, adaptability, dependability, sustainability, and quality of what is produced over time, all of which is influenced by factors such as resources, working conditions, developer experience, and sociocognitive factors that drive effective problem-solving and developer wellbeing, such as the Developer Thriving Framework [7, 9-13]. In the context of engineering teams, performance may be tracked by calculating DORA metrics [14], examining code churn, lead times, longevity, and dependability, and finding signals for the factors outlined in the Developer Thriving Framework, such as teams that prioritize learning and correcting maladaptive patterns over non-stop production, or teams that collaboratively define and measure success [10]. Performance definitions create the ability to look at team output, while wrapping in appropriate and helpful context, thinking about long-term and cross-team impacts, and considering the systems in place to create an environment conducive to sustainable high performance - all of which could provide protection against external frictions or constraints that create low "production" moments.

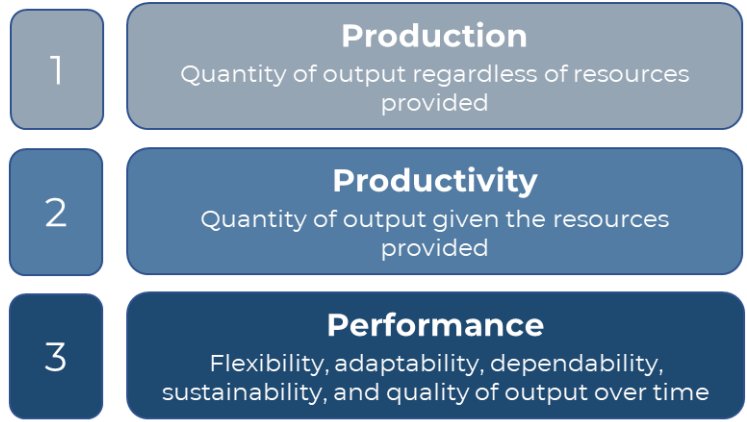

Fig 1. Three Layer Productivity Framework

**Present Study**
Given the distinct meanings of production, productivity, and performance, it may be helpful for managers and leaders to think about which they have available to learn from and are aiming to improve, rather than simply mislabeling all software metrics as "productivity" metrics. Failing to do so could create confusion across software teams about how their work is evaluated.

Using the Three Layer Productivity Framework as a guide, we conducted an intercept study to examine which metrics individual contributors, managers, and engineering leaders believe their organizations are currently using. We also examined which metrics individuals preferred their organizations used to examine how adjustments could be made to better capture developers' work. Finally, we examined the differences in current versus preferred use by roles (engineering leader, manager, and individual contributor). The aim of our exploratory pilot study was to demonstrate how the Three Layer Framework (without naming it as such) resonated with software engineers, and helped to explain these observed differences between what these engineers *say* their organizations see as productive, versus what engineers *wish* their organizations would move towards.

**METHODS**
**Participants & Recruitment**
An intercept study focuses on recruiting participants in places they would normally frequent and is a type of qualitative field research which focuses on high external validity (e.g., collecting data with participants in the course of their real experience). We recruited 16 participants using convenience sampling at the LeadDev 2023 Conference in New York.

We started by announcing our presence at a central booth in the middle of the conference. Participants approached our booth at will and completed a short (5-7 minute) experimental session with a researcher. All participants gave verbal assent to the study and were informed of the Developer Success Lab's consent & participant privacy policies (Appendix D). We provided mint candy as incentives to all participants. A summary of participant demographic and firmographic characteristics can be seen in Table 1.

| Characteristic | n |
| --- | --- |
| N = 16 | |
| Gender | |
|    Man | 13 |
|    Woman | 2 |
|    Prefer not to answer | 1 |
| Role | |
|    Leader | 5 |
|    Manager | 7 |
|    Individual Contributor | 5 |
| Industry | |
|    Financial Services | 5 |
|    Industrials/Manufacturing | 5 |
|    Media/Entertainment | 1 |
|    Technology | 1 |
|    Other | 2 |
| Race/Ethnicity | |
|    Black/African American | 1 |
|    East Asian | 1 |
|    Latinx | 1 |
|    Middle Eastern/ North African | 1 |
|    South/Southeast Asian | 1 |
|    White | 10 |
|    Prefer not to respond | 1 |

Table 1. Participant Characteristics

**Methodology & Measurement**

Because we were interested in evaluating developers' mental models on the difference between production, productivity and performance, we utilized the **Card Sort methodology** taken from user experience and product research fields [15]. In a Card Sort, participants organize cards containing different items into affinity groups. This method is used to observe how participants intuitively group and label information.

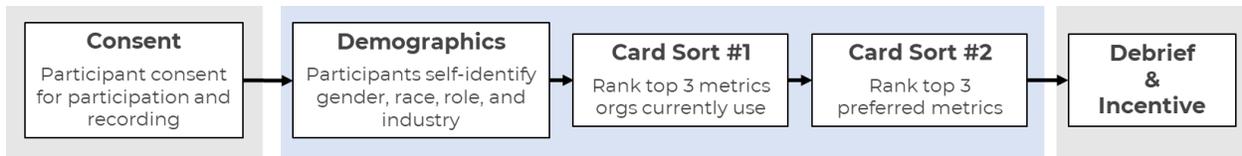

Fig 2. Intercept Study Flow

To do this, we created a list of terms associated with the developer workflow cycle and categorized these words into "production", "productivity" and "performance" based on prior literature definitions. For example, our production and performance terms were taken from previous research categorizing common software metrics in terms of raw activity and performance [2], and our productivity terms were taken from the Three Layer Productivity framework [6]. After generating a list of terms, we created a final list of three production terms, three productivity terms, and four performance terms (see Table 2).

| Production | Productivity | Performance |
|---|---|---|
| ● # of code commits<br>● # of pull requests<br>● # of code reviews | ● Team size<br>● Tech debt size<br>● Available external tools | ● Code quality<br>● Code longevity<br>● Usability and value of work to the org<br>● Code review quality |

Table 2. Final production, productivity, and performance terms

To gather data, we included each term onto a separate index card in preparation for the intercept study. We then developed a script asking participants to complete two main card sorting tasks: 1) to rank the top 3 ways they believe developer productivity is currently being measured in their organization, and 2) rank the top 3 ways that they would prefer developer productivity to be measured in their organization (see Figure 3; see Appendix A for script).

During the conference, we laid out each of the cards on a table, making sure to randomize the cards order so that the participant was not primed or presented with pre-defined categories. The participants then selected their preferred cards based on the questions asked from the script.

Qualitative field studies such as intercept studies are typically conducted as very brief sessions, so as to not disrupt participants' experiences and to encourage broader participation by making the experience easy for participants. Each study ran approximately 5-7 minutes. Participants' responses were documented in a Google Form by a researcher and recorded by a smartphone video clamped to the table. The video only recorded the card sort to observe participants' notable motor behavior (i.e. delays, pauses) in a similar fashion to an eye-tracking system (see Figure 4). The video did not record participants' faces. Participants provided consent for recording and were able to complete the study without recording if desired. All recordings were stored onto a secure server accessible only to the study's researchers. Although a phone recording system cannot be held at the same weight as an eye-tracking system, it is still an accepted method of observing participants' motor behavior and is a lightweight way to include meaningful information such as participant hesitation. Although there were no noticeable differences in hesitation between questions or in choosing particular metrics, most participants took longer to rank their preferred metrics than their current metrics.

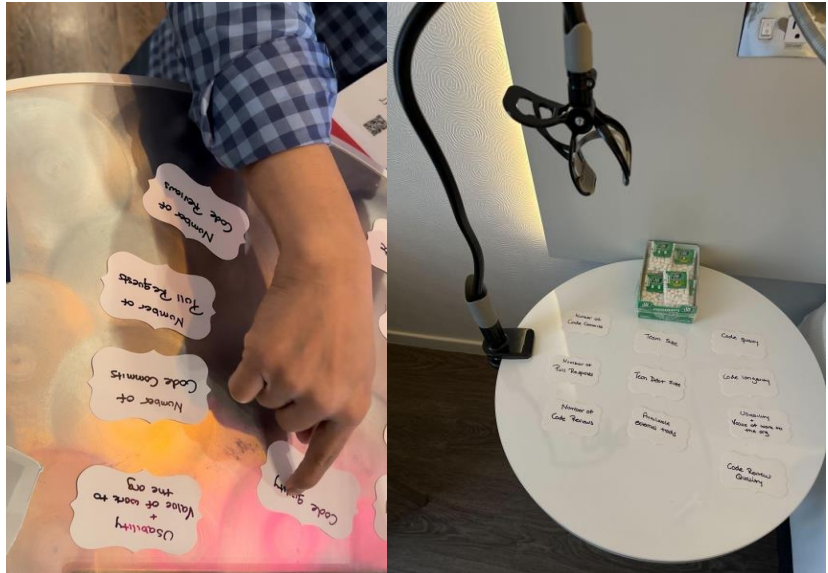

Fig 3. Running the Intercept study at the LeadDev New York 2023 Conference

**RESULTS**

To interpret participants' rankings, we calculated weighted values based on participants' rankings of each metric (ranked scoring method). Under this calculation, metrics ranked higher were assigned greater values or "weights." Specifically, metrics ranked first were assigned a value of 3, metrics ranked second were assigned a value of 2, and metrics ranked third were assigned a value of 1. Unranked metrics were assigned a value of zero. This is a typical method of interpreting ranked data in the social sciences [16].

Our results showed there were differences between how participants think organizations currently measure and how they prefer organizations measure developers' work. Specifically, participants cited usability and value of work to the organization, followed by code longevity, code quality, and number of pull requests, as the top metrics currently used to measure developers' work. However, participants reported that they preferred organizations to prioritize code quality, followed by usability and value of work to the organization and code longevity (For a breakdown of each metric's ranking distribution see Appendix B).

The findings also revealed differences in reported current and preferred metrics use; **the most notable discrepancies** were found for code quality, number of code commits, and number of pull requests. Here, participants reported a strong preference for increased use of code quality metrics than is currently used, and a decreased use of metrics around the number of code commits and pull requests than is currently used (see Figure 4). Of note, **zero** participants selected number of code commits and number of code reviews as their preferred metrics, despite some reporting organizational use of these metrics.

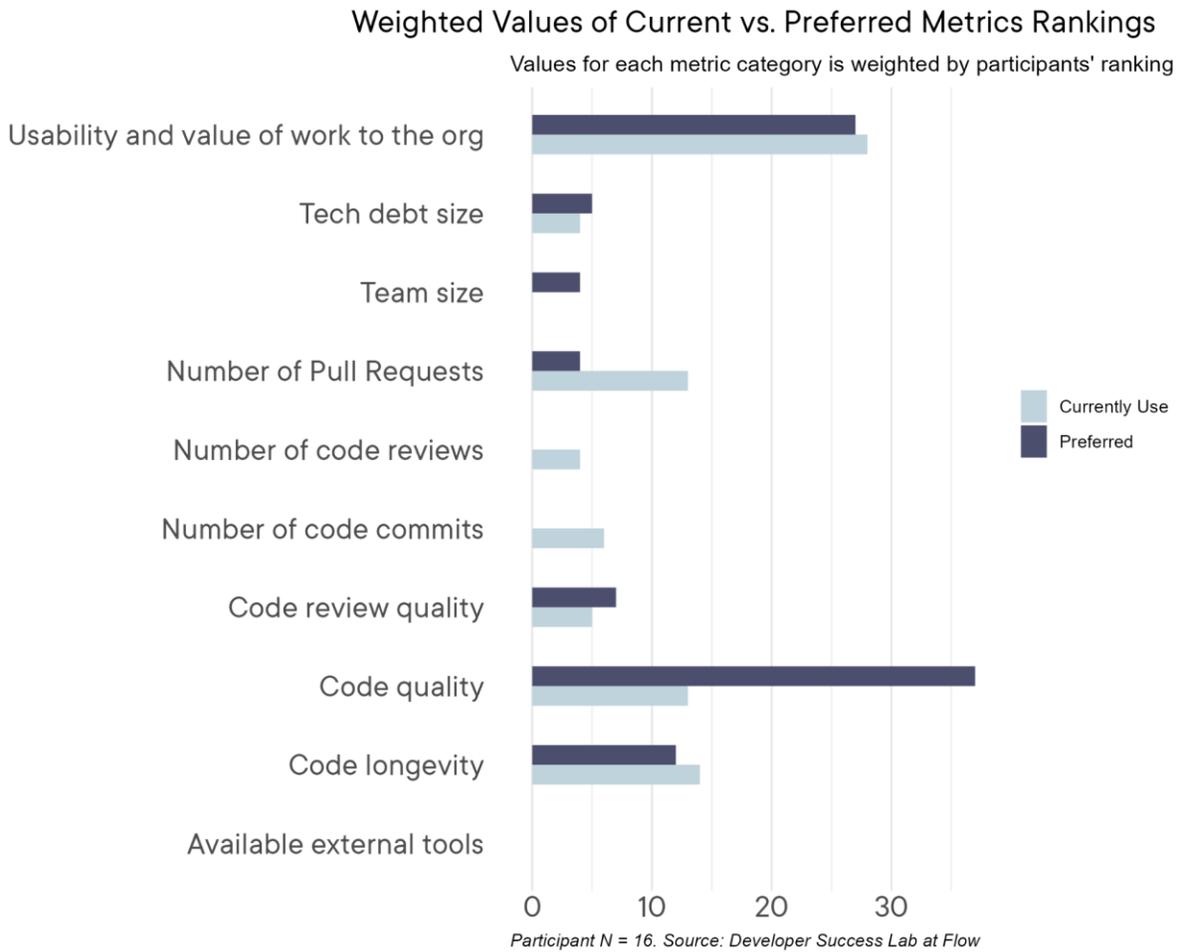

Fig 4. Differences in reported current and preferred metrics used in all participants.

When we aggregated and categorized these metrics within the Three Layer Productivity Framework, our results suggested that these discrepancies could be traced back to differences in the approach participants believe their organizations take toward metrics. Specifically, participants reported a current focus on performance, followed by production and productivity layer metrics, but participants ranked their own preference as performance, followed by productivity and production layer metrics. Participants also showed an overall preference for an increased focus on productivity and performance metrics, but a decreased focus on production layer metrics (see Figure 5).

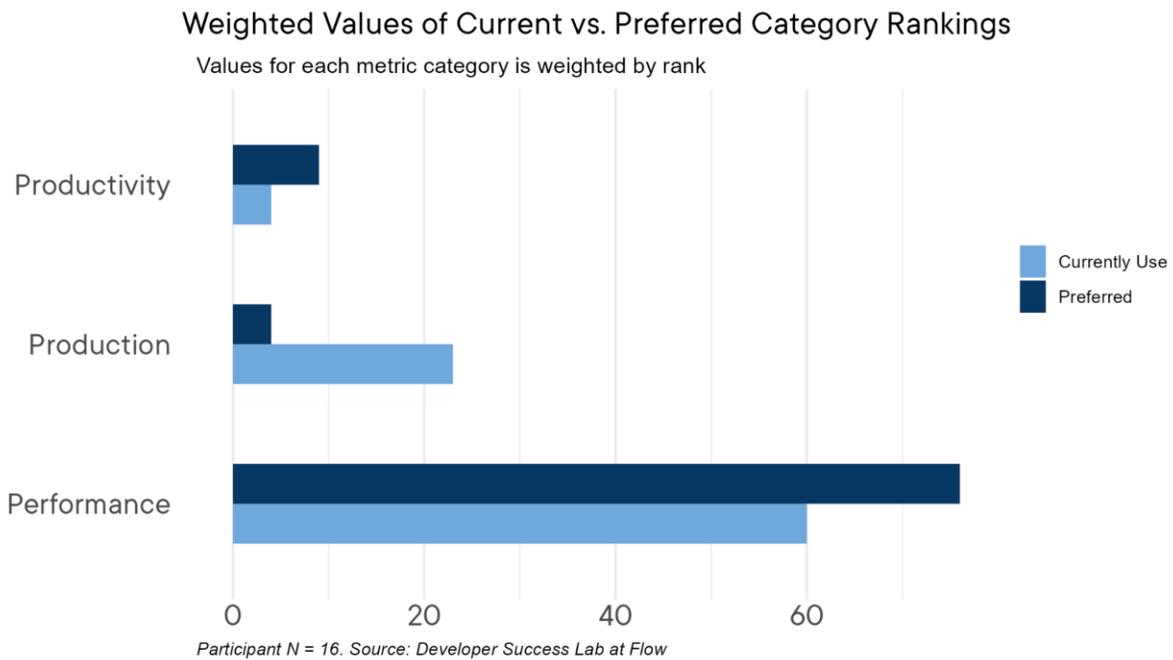

Fig 5. Differences in reported current and preferred metrics categories used in all participants.

Taken together, the findings indicated that our participants largely preferred if organizations continued to approach measurement using performance layer metrics such as code quality, usability and value of work to the org, and code longevity. In contrast, our participants reported a preference for a decreased focus on production layer metrics such as raw counts of code commits and pull requests, and an increased focus on productivity layer metrics such as the consideration of contexts like team size or tech debt.

Within this relatively small case study, it is important to note that any subgroup differences between leaders, managers, and individual contributors are not intended to be representative evidence about how these groups feel at large. However, we explored the distribution of responses across these roles as an interesting signal for potential differences that future research should investigate more deeply. Our findings highlighted that managers see the productivity landscape differently than either leaders or individual contributors.

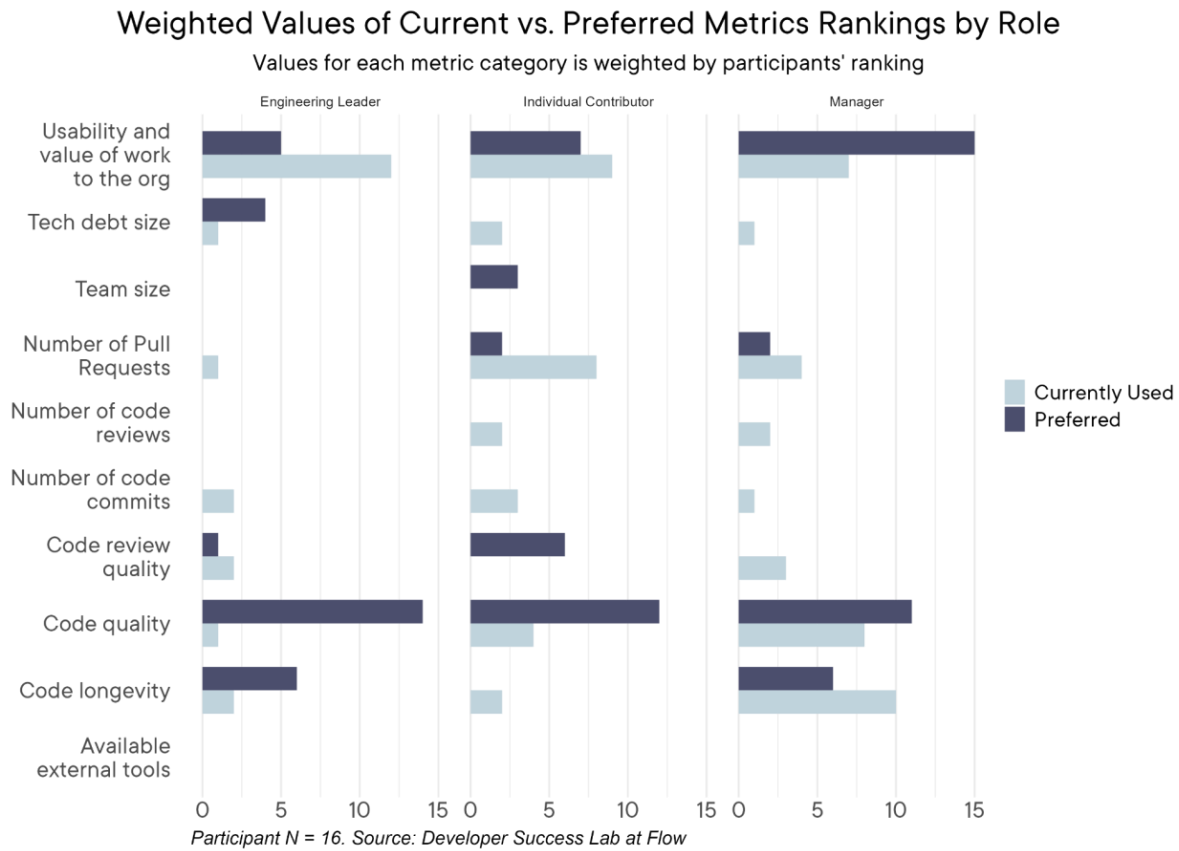

Fig 6. Differences in reported current and preferred metrics used by role.

Figure 6 summarizes participants' weighted rankings of these metrics by role. Observationally, both engineering leaders and ICs ranked usability and value of work to the org as the top definitions currently used, but indicated their biggest preference was for organizations to shift to prioritizing code quality to measure developers' work. However, this pattern was flipped for managers: managers reported that code longevity, followed by code quality, were the primary metrics currently used, but indicated their biggest preference was for organizations to shift to usability and value of work to the org. While all groups emphasized code quality in their preference, managers' desire for their orgs to shift to **usability and value to the org** was larger than the other groups.

Interestingly, individual contributors uniquely ranked the **number of pull requests** as a strong current focus of their organizations, and reported a moderate preference for prioritizing code review quality - a perspective that engineering leaders and managers did not share. Individual contributors were also unique in calling for an organizational consideration of "team size." For a breakdown of each ranking distribution by role see Appendix C.

When we aggregated these metrics with the Three Layer Productivity Framework and examined them by role, our results suggested that these discrepancies could again be traced back to perceived differences in the approach organizations reportedly take toward metrics by role. Across roles, participants reported a strong current and preferred focus on performance layer metrics - a focus they also wanted their organizations to increase. However, compared to the other roles, ICs reported a stronger current focus on production layer metrics, a focus that all roles preferred would decrease. ICs were also more likely to report a current focus on productivity layer metrics, a focus that they and engineering leaders preferred would increase. (see Figure 7).

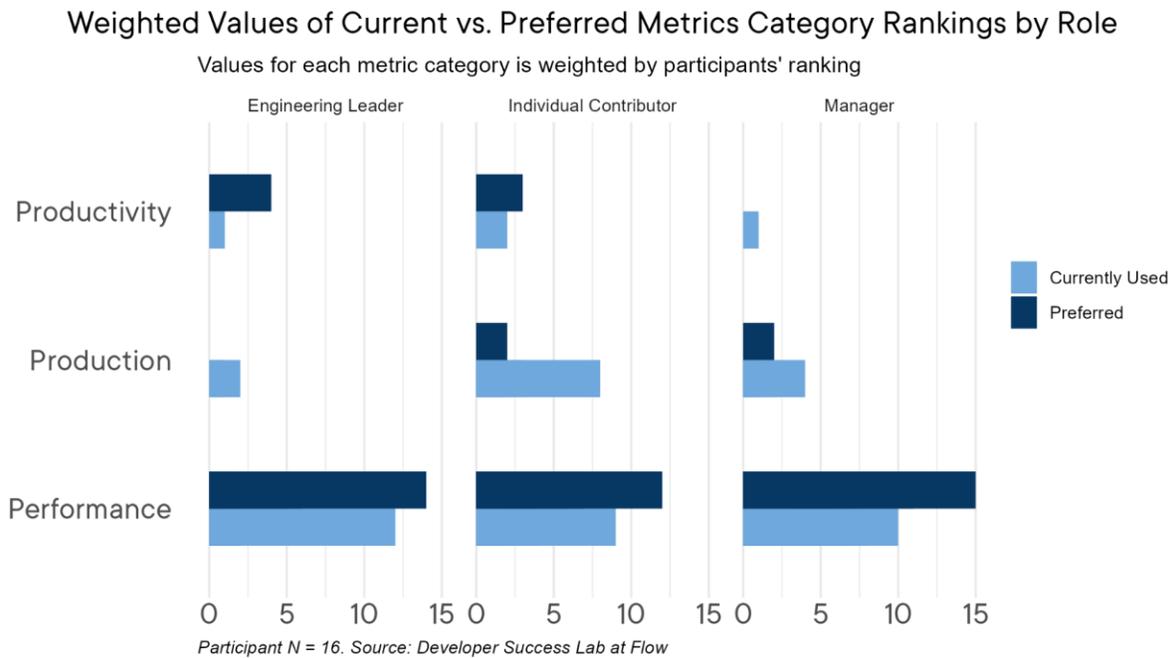

Fig 7. Differences in reported current and preferred metrics categories by role.

**DISCUSSION**

In this report, we present the Three Layer Productivity Framework as a conceptual step towards triangulating *why* software teams might be struggling with conflicts over productivity definitions. In a targeted case study, we show a model for an exploratory "first step" in better understanding developers' thoughts on how their productivity is currently instrumented in their organizations, versus how they want their productivity to be understood. This example case study could be replicated within moments where implicit organizational mental models of "productivity" are likely to come into conflict. For example, we believe that sorting "software metrics" into the Three Layer Productivity Framework could be a useful exercise for groups as diverse as a team of senior engineering leaders setting organizational targets, as a reflection exercise for engineering bootcamps or hackathons, or a mentorship exercise for junior developers.

Even in this case study, we saw signals of misalignment on the definitions of "productivity" and success for developers and their organizations (see Figure 8). Overall, our case study also found that a research backed Three Layer Productivity Framework (aggregating types of metrics or measurement into production, productivity, and performance) was a thought-provoking framing for ICs, managers, and engineering leaders. While organizations and individuals both sought to think about "performance" in software work, individuals reported an organizational emphasis on production that they themselves disagreed with. Our findings suggest there are discernible differences between current state and ideal state, and in particular, a diagnosis that organizations need to shift away from production and towards performance. This is particularly striking given that many of our participants were themselves engineering leaders and managers. Future work should more deeply explore these misalignments and how engineering organizations can learn from them to transform their practices.

Although this case study provides exploratory data towards this question, and more research is needed to understand engineers' perceptions at scale, our findings echo evidence of a misalignment in how engineers and their organizations define, and thus measure, productivity that has been found in large-scale developer research [2]. This points to a need for engineering organizations to prioritize communication and collaborative decision making on how developers' work is measured both within organizations and across the industry. This is particularly important when considering that collaboratively defining success increases developer agency, a known factor of developer thriving and predictor of software team performance [10]. These differences represent an especially important opportunity gap for businesses engaged in transforming their engineering organizations. Developers' insights about what they wish their organizations would measure may be the key to moving towards better, and more actionable sources of evidence about engineering work.

While all participants were aligned on performance metrics being a top preference, there was misalignment on which specific metrics were used and preferred. For example, relative to the other roles, ICs were more likely to report a current focus on production layer metrics such as pull requests and code commits and were more likely to cite team sizes and the quality of code reviews as preferred measures for developers' work. This may reflect two things: 1) the relative ease and accessibility of tracking individual work, which production metrics do without the need for team or organization-level contexts and 2) the greater influence that engineering teammates have on ICs in reducing personal friction and completing work - thus making metrics related to team size and code review quality more valued.

In contrast, relative to the other roles, managers were more likely to report a preferred focus on the usability and value of work to the org and engineering leaders were more likely to report a preference for metrics around code quality. This may reflect the nature of management roles, which require not only tracking work within a team, but also tracking work between teams and within the larger organization. For managers, usability and value of the work to the org may feel particularly salient due to a responsibility for making their team's work visible and aligned with organizational goals. For engineering leaders, code quality may feel particularly salient due to a responsibility for the long-term and sustainable success of engineering teams, codebases, and systems.

Notably, we also observed that no participants reported "available external tools" as either a current organizational metric for developer work, or as a preference to be considered when thinking about their productivity. This may reflect a sample of engineers who feel they are already supplied with all the tools they need, but it also suggests that developers do not think about "tools" as their first source of evidence for developer productivity in a business. This contrasts with many industry-wide discussions of developer experience that center "experience" almost entirely on the availability and quality of developer tooling [17].

In sum, our results suggest that organizations should continue to approach measurement through the performance layer of productivity by not only measuring usability and value of work to the org, but also integrating measures of code quality, code review quality, and code longevity. In contrast, our results suggest that organizations should decrease their focus on production metrics like the number of pull requests, and eliminate raw counts of code commits and code reviews. Finally, organizations could consider integrating productivity metrics by accounting for team and organizational resources and by potentially looking into contextual variables such as team size and tech debt (see Table 3).

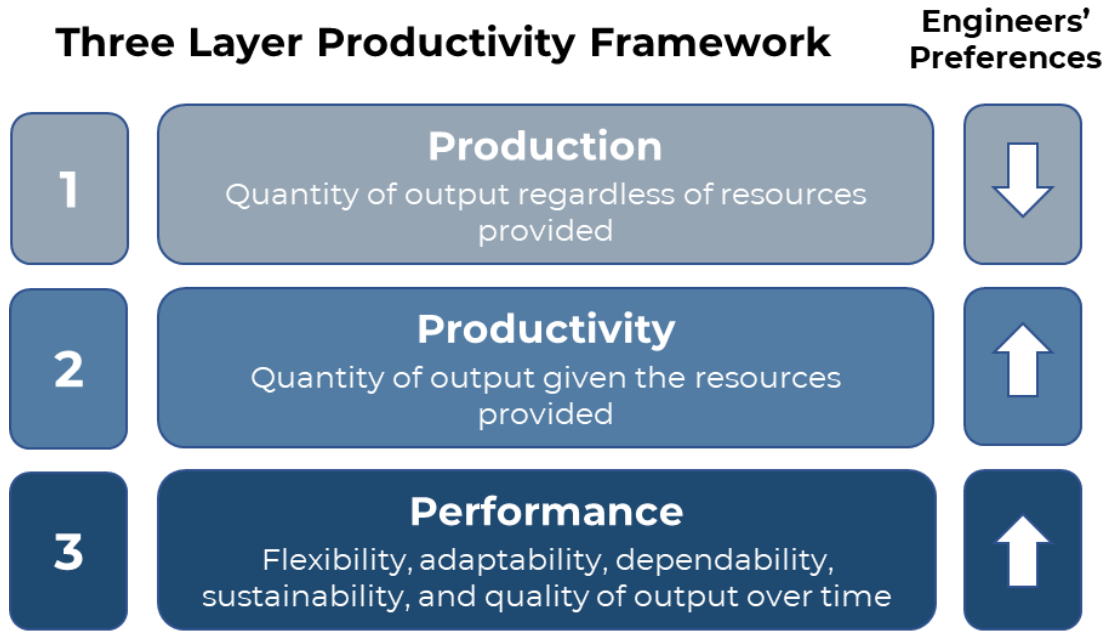

Fig 8. Three Layer Productivity Framework and preferences for increasing/decreasing its emphasis inside of engineering organizations.

| Finding | Insight | Recommendations | Potential impact |
| --- | --- | --- | --- |
| Across roles (leaders, ICs, Managers), performance layer metrics are preferred over production and productivity layer metrics. | There is **alignment** on measuring performance | Organizations should **continue** using performance layer metrics | ⬆ Thriving across roles <br> ⬆ Retention across roles <br> ⬆ Accuracy in assessing progress toward org targets <br> ⬆ Reliable value delivery to customers <br> ⬆ Long-term and resilient cycles of innovation |
| Participants report currently using productivity layer metrics less than preferred | There is **misalignment** measuring productivity | Organizations should consider an **increased** focus on productivity layer metrics | ⬆ Clarifies expectations <br><br> ⬆ Ability to identify external friction and constraints <br><br> ⬆ Accuracy in comparing teams <br> ⬆ Thriving across roles, particularly ICs <br> ⬇ Time to address external friction and constraints like tech debt |
| Participants report currently using production layer metics more than preferred | There is **misalignment** measuring production | Organizations should consider a **decreased** focus on production layer metrics | ⬆ Thriving across roles, particularly ICs <br> ⬆ Retention across roles <br> ⬇ Inaccurate comparisons between teams <br><br> ⬇ Inaccurate assessments of external friction and constraints like tech debt |

Table 3. Summary of Key Insights and Recommendations

APPENDIX A
**Intercept Study Script**

| Category | Questions |
| --- | --- |
| Demographics | Which category below best describes your current role? How many years of experience do you have? What is the principal industry of your organization? (Optional) How do you identify with regard to race/ethnicity? (Optional) How do you identify with regard to gender identity? |
| Ranking | Using the words here, rank the top 3 ways that you think developer productivity is currently being measured at your organization. Using the words here, rank the top 3 ways that you wish developer productivity would be measured at your organization. |
| Closing | If you would like to join our lab mailing list please enter your email address below. We'll use this to very occasionally contact folks for future studies (~2 times a year) and you'll receive the results of this study directly: |

# APPENDIX B: Breakdown of Rankings in All Participants

**Top Ranked Metrics**

The majority of participants ranked usability and value of work to the org as their organization's current top method of measuring developers' work (60%), a focus that 50% of our participants preferred. However, 20% of participants instead reported a current focus on the number of pull requests, a primary focus that none of our participants preferred. Notably, though only 6.67% of participants reported a current primary focus on code quality, 44% ranked code quality as their preferred top metric.

**Second Ranked Metrics**

The importance of code quality was further highlighted when examining the metrics ranked second. Although participants were equally split between code quality and longevity as organizations' current secondary focus (28.57% each), the majority of participants preferred a secondary focus on code quality (43.75%) over code longevity (25%).

**Third Ranked Metrics**

When examining the metrics ranked third, the results highlighted the greatest amount of non-performance metrics represented. The majority of participants ranked the number of pull requests (28.57%) as their organization's tertiary focus, a focus that only 12.5% of participants preferred. Instead, most participants reported wanting a tertiary focus on code longevity (25%), code review quality (18.25%), or output in the context of tech debt size (18.75%) instead. Although a moderate proportion of participants reported a current tertiary focus on code longevity (25%), none reported a current tertiary focus on code review quality.

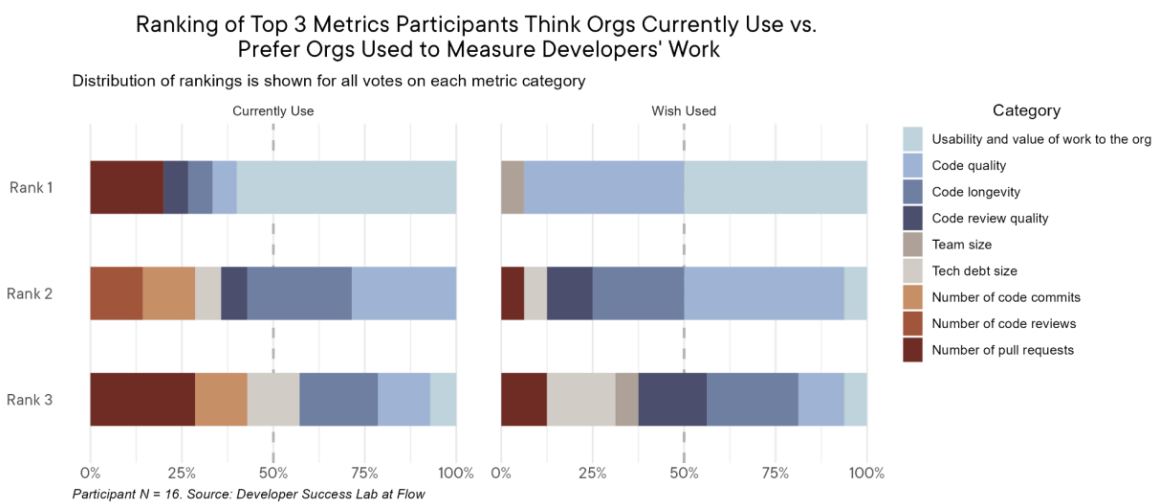

Fig 9. Distribution of participant rankings of current vs. preferred metrics organizations use.

**Top Ranked Metrics Category**
When categorizing these metrics within the Three Layer Productivity Framework, our findings showed that 80% of participants highlighted performance metrics as their organization's current priority, a focus that 94% also preferred. However, 20% of participants also reported that their organizations were prioritizing production metrics - a focus that none preferred. Interestingly, none reported a current focus on productivity metrics.

**Second and Third Ranked Metrics Category**
Looking at the secondary and tertiary priorities, these overall trends continued, with the majority of participants citing performance metrics as their current and preferred focuses, though to lesser degrees. Instead, an increasing proportion of participants cited a current focus on production metrics, a focus that most preferred to replace with performance or productivity metrics. For example, although 42.86% of participants cited production metrics and 14.28% cited productivity metrics as their organization's tertiary focus, only 12.5% preferred production metrics and the proportion of participants preferring productivity metrics went up to 25%.

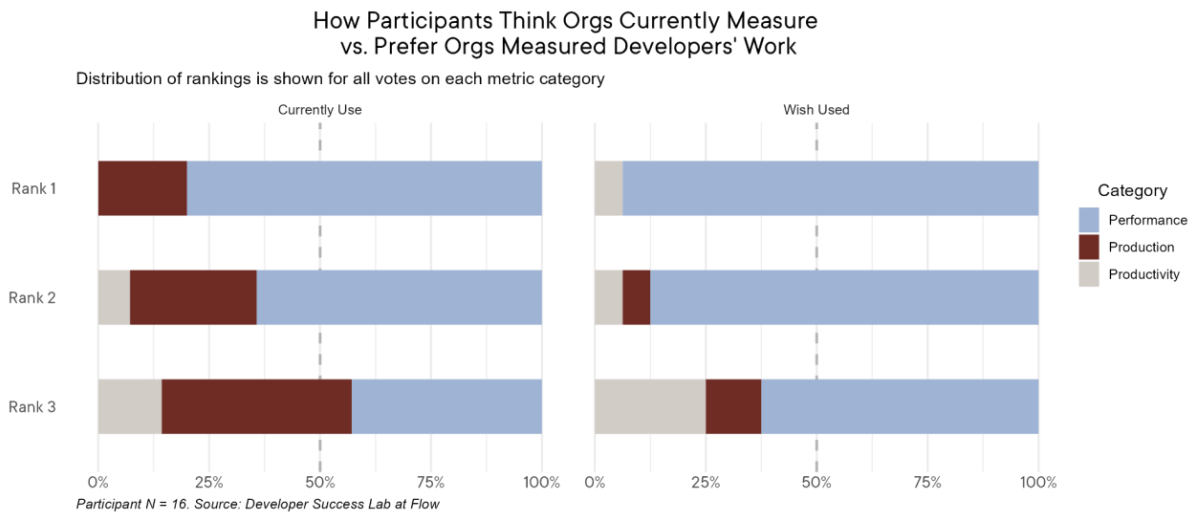

Fig 10. Distribution of participant rankings of current vs. preferred metrics categories organizations use.

**APPENDIX C. Breakdown of Rankings by Role**

**Top Ranked Metrics Category by Role**
Across roles, the majority of participants reported a current focus on the usability and value of work to the org. However, while 100% of engineering leaders reported that their organization prioritized the usability and value of work to the org, only 20% reported this as a preference. Instead, the majority of engineering leaders reported a strong preference for code quality (80%). In contrast, though only 33% of managers reported that their organization prioritized the usability and value of work to the org, 83% preferred a primary focus on the usability and value of work the org.

Interestingly, ICs were fairly split on their respective organization's current focus, with 60% of ICs citing usability and value of work to the org and 40% citing a focus on the number of pull requests. Notably, none preferred a continued focus on pull requests, instead citing a preference for a focus on usability and value of work to the org or code quality (40% each).

**Second Ranked Metrics by Role**
The results highlighted a greater focus on code quality and longevity when comparing organizations' secondary focuses by role. Although our engineering leaders were equally split between code longevity, code review quality, and the number of code commits as their organization's current secondary focus (33% each), none of them preferred a secondary focus on code review quality or the number of code commits. Instead, the majority of engineering leaders preferred a secondary focus on code longevity (40%).

Interestingly, 50% of our managers reported a current secondary focus on code longevity. However, unlike our engineering leaders, the majority instead reported a preference for a secondary focus on code quality (50%). In contrast, none of our ICs reported a current secondary focus on code review quality or code longevity, instead citing a current secondary focus on code quality (40%). Additionally, they were fairly split between preferring a secondary focus on code quality (60%) and code review quality (40%).

**Third Ranked Metrics by Role**
We had the lowest within-group consensus on organizations' current tertiary priorities. Engineering leaders were equally split between number of pull requests, tech debt size, and code quality (33% each). Of these metrics, only tech debt size was a preferred tertiary focus, along with the addition of code longevity (40% each). In contrast, most ICs were split between code longevity and number of pull requests (40% each). Unlike engineering leaders, ICs did not wish for a tertiary focus on longevity, instead preferring a tertiary focus on the number of pull requests and code review quality (40% each).

Strikingly, not a single manager agreed on their respective organizations' current tertiary focus, resulting in an equal split between usability and value of work to the org, code quality, code longevity, tech debt, number of commits, and number of pull requests. Similar to our engineering leaders though, the majority of our managers preferred quality and longevity as their desired tertiary focus (33.33% each).

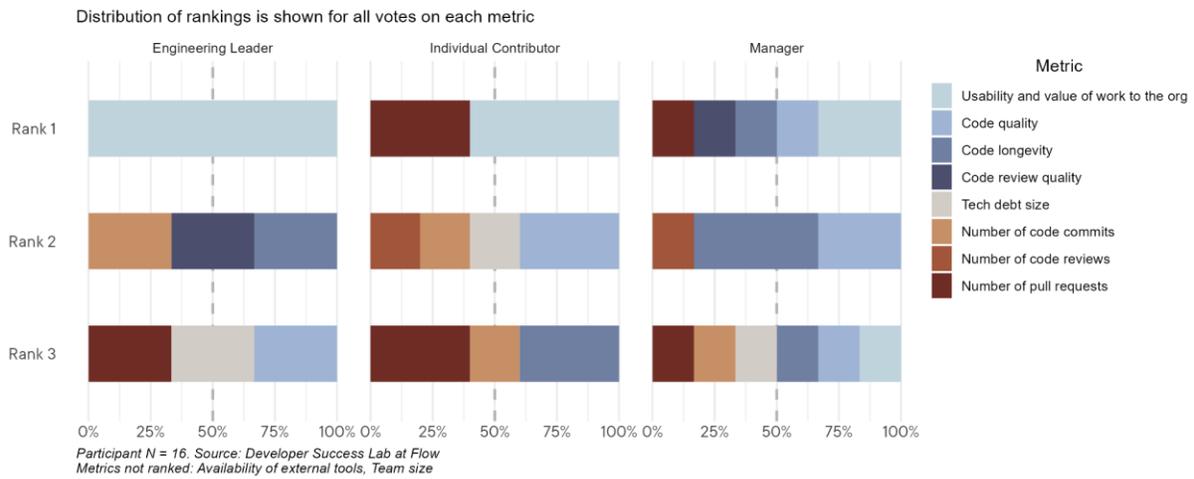

Fig 11. Distribution of participant rankings of metrics organizations currently use by role.

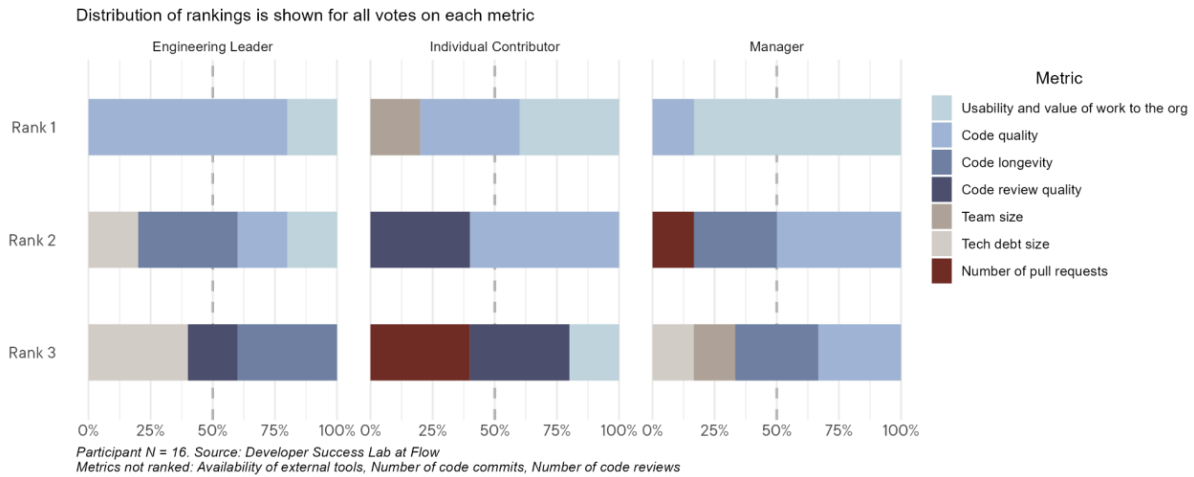

Fig 12. Distribution of rankings of metrics participants prefer their organizations use by role.

**Top Ranked Metrics Category by Role**

When categorizing these metrics within the Three Layer Productivity Framework, the results indicated that the majority of participants across roles reported a *current* focus on performance metrics - a focus that the vast majority also preferred. However, 40% of ICs reported that their organization instead prioritized production metrics. In comparison, only 17% of managers and none of our engineering leaders reported a primary focus on production metrics. Interestingly, none of our participants reported a current focus on productivity metrics and only 20% of ICs preferred this as a primary focus.

**Second and Third Ranked Metrics Category by Role**

The secondary and tertiary priority rankings highlighted a similar trend, with engineering leaders and managers typically reporting a current focus on performance metrics and ICs continuing to report an equal or greater focus on production metrics. Notably, none of our engineering leaders expressed a preference for a secondary or tertiary focus on production metrics. In contrast, ICs were fairly split between performance (60%) and production (40%) metrics as a preferred tertiary focus.

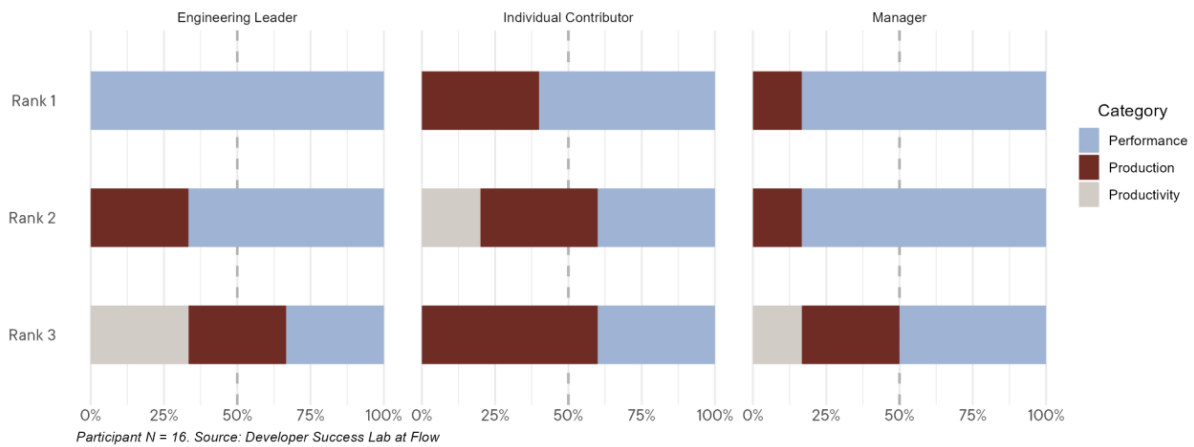

Fig 13. Distribution of participant rankings of metrics categories organizations currently use by role.

.

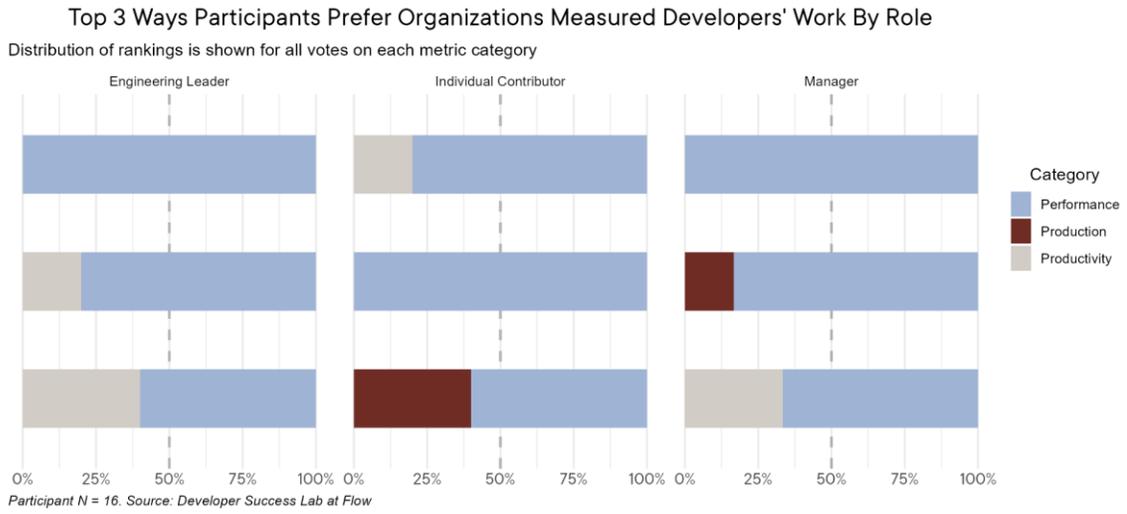

Fig 14. Distribution of rankings of metrics categories participants prefer their organizations use by role

**APPENDIX D**
**Sample Participant Consent Form**

*Developer Success Lab*
*This survey is from the Developer Success Lab at Flow. We are interested in learning more about developers at work. This survey is a part of our ongoing research. By participating in this survey, you are giving consent for us to use your survey responses in our ongoing research, including public reports, white papers, and conference presentations. Please read this consent form carefully to ensure you understand how your data will be used.*

*It is important to know that there are no wrong or right answers on this survey. Our research team is interested in hearing your most authentic answers to any question.*

*Participant data & analysis*
*The Developer Success Lab cares deeply about participant privacy. Your responses on this survey will be analyzed only by our researchers. In order to learn from this data, researchers will have access to everything that you share on this survey. Please do not share anything you do not wish to share with our research team.*

*Findings*
*Our research is communicated to an external audience as part of Flow's contributions to the tech community, where we believe research findings can benefit developers and their teams (for example, conference presentations, public white papers). When we share research, all findings will be anonymized, removing any names or specific team contexts. Quantitative insights will be summarized in aggregate as part of a large research report. For example, we may share statements like: "10% of respondents agreed that..."*

*Any quotes and text responses that you give as a part of this research may also be shared in our external reports and may be quoted directly. We will anonymize all quotes, removing any specific mention of teams, names, or contexts that might be personally identifiable. Where necessary, we may provide anonymized contextual details (for example, "a senior engineer working primarily on backend team noted that....").*

*Questions, Concerns, and Opt-out.*
*While we hope that you will enjoy completing this survey and sharing your insights with us, you may choose to leave this survey at any time. You may also leave any questions blank that you do not wish to answer.*

*If you have any questions or concerns you can reach out to the Developer Success Lab directly [EMAIL]*